\documentclass[aps,prb,twocolumn,superscriptaddress,10pt]{revtex4-2}
\usepackage{graphicx} 
\usepackage{amsmath} 
\usepackage{amssymb} 
\usepackage{bbold}
\usepackage{enumitem}
\usepackage[T1]{fontenc}
\usepackage[utf8]{inputenc}
\usepackage[urlcolor=blue,colorlinks=true,citecolor=blue,linkcolor=blue,pdfstartview={FitH},bookmarks=false]{hyperref}
\usepackage{xcolor}

\begin{document}
\title{Charge order on a triangular lattice with Mott physics and arbitrary charge density}

\author{Aleksey Alekseev}
\homepage[\mbox{ORCID ID}: ]{https://orcid.org/0000-0001-5102-6647}
\affiliation{\mbox{Institute of Spintronics and Quantum Information, Faculty of Physics and Astronomy}, Adam Mickiewicz University in Poznań, Uniwersytetu Poznańskiego 2, PL-61614 Poznań, Poland}
\author{Agnieszka Cichy}
\homepage[\mbox{ORCID ID}: ]{https://orcid.org/0000-0001-5835-9807}
\affiliation{\mbox{Institute of Spintronics and Quantum Information, Faculty of Physics and Astronomy}, Adam Mickiewicz University in Poznań, Uniwersytetu Poznańskiego 2, PL-61614 Poznań, Poland}
\affiliation{Institut f\"ur Physik, Johannes Gutenberg-Universit\"at Mainz, Staudingerweg 9, D-55099 Mainz, Germany}
\author{Konrad Jerzy Kapcia}
\email[Contact author: ]{konrad.kapcia@amu.edu.pl}
\homepage[\mbox{ORCID ID}: ]{https://orcid.org/0000-0001-8842-1886}
\affiliation{\mbox{Institute of Spintronics and Quantum Information, Faculty of Physics and Astronomy}, Adam Mickiewicz University in Poznań, Uniwersytetu Poznańskiego 2, PL-61614 Poznań, Poland}

\date{\today}
 
\begin{abstract}
Triangular-lattice systems attract a lot of attention due to various frustration-induced and strongly correlated effects. Here, we focus on the charge-ordering phenomenon by means of investigation of the extended Hubbard model with dynamical mean-field theory (DMFT). By considering the intersite nearest-neighbor interaction we have found a very rich phase diagram that contains large number of features, phases, and phase transitions. Among them are pinball-liquid (PL) phases where we distinguish charge-transfer-driven and Mott-localization-driven PLs; phase transitions that change their order as model parameters evolve (from discontinuous to continuous); very strong particle-hole asymmetry. Various features of the phase diagram are found to be better understood by means of the simple mean-field approximation (MFA). Moreover, besides helping with interpretation of the phase diagram, the MFA results together with results for the atomic-limit model are found to be able to set rather good expectations on how the DMFT phase diagram should look like. Nevertheless, a few features were not expected and are found within the DMFT, such as a small-region intermediate metallic phase on an electron-doped side of the phase diagram.
\end{abstract}

\maketitle

\section{Introduction}

A variety of materials have a structure of a triangular lattice, such as organic conductors, transition-metal oxides and dichalcogenides, or even $^3$He atoms adsorbed on a surface. Particularly attractive are the moir\'e lattices \cite{kumar2025origin, ZhaiRPP2025, xia2026bandwidth, han2026topological} that can also be described by the triangular-lattice model \cite{Wu2018,NuckollsNRM2024,MakNSR2026} while their interaction and hopping parameters can be varied with the change of a twisting angle, layer displacement, and by choosing 2D layers from a rich family of 2D materials.
The interest behind their investigation comes from the fact that they exhibit such phenomena as superconductivity, various exotic magnetic and charge orderings \cite{XuPRB2024,gomez2026interleaved}, topological states and else, especially those that are frustration-induced \cite{AndersonMRB1973,BalentsNature2010,saha2025polarization, gomez2026interleaved} and strongly correlated ones. 
Here, we focus on the charge-ordering phenomenon to reveal the rich capacity of frustrated geometries with Mott physics to form charge-ordered phases with complex relations between each other and with non-charge-ordered phases. 
Note that the charge ordering interplays or competes with the superconducting phases which makes its understanding desirable \cite{CaoNat2018A,CaoNat2018B}. 
Among the charge-ordered phases researchers distinguish Wigner crystals, generalized Wigner crystals \cite{WignerPR1934,LenacPRB1995,kumar2025origin,wang2025intrinsic}, charge-density waves, charge-transfer insulators. Such a separation is not essential within the framework of our work.
In the context of the paper, a pinball liquid \cite{hotta2006, hotta2007, Merino2013, Ralko2015, trousselet2012valence, miyazaki2009variational} worth highlighting and defining. Such a charge-ordered phase is commensurate with a $\sqrt{3}\times\sqrt{3}$ supercell with three sublattices where charge carriers on sites of one sublattice (pins of a pinball liquid) are localized, while two other sublattices form a honeycomb lattice which is a conducting medium. One may compare the pinball liquid with supersolids that has itinerant and localized particles as well. 

To investigate the charge order on the triangular lattice we solve the extended Hubbard model (EHM) with nearest-neighbor (NN) density-density interaction on a triangular lattice within the dynamical mean-field theory (DMFT) \cite{GeorgesRMP1996,ImadaRMP1998}. We do not limit ourselves to any particular material, thus building the knowledge of what is generally possible and what the mechanism behind various phase transitions and phases is. It is especially relevant considering that the moir\'e lattices are highly tunable, and additionally, considering that the EHM is now not just models a system but explicitly describes it in the context of ultracold atomic gases on optical lattices \cite{Wessel2007, Mathey2007, Struck2011, Jo2012, Yamamoto2020, yang2021, Mongkolkiattichai2023, Wu2024}. Note also that the EHM is the most natural extension of the simple, artificial, but extremely successful Hubbard model \cite{Hubbard1963,PennPR1966} that provided a lot of insights into the strongly correlated physics.

We consider the EHM on a triangular lattice within the grand canonical ensemble with an all-encompassing range of chemical potentials without fixing the total charge density to any particular value. 
The work is motivated by the fact that such a whole picture is, to our knowledge, not available in the literature.
Meanwhile, there are results with fixed filling of the lattice, such as quarter filling for organic conductors \cite{seo2000, Calandra2002, kaneko2006, Merino2013}, and a variety filling factors for the moir\'e lattices \cite{Pan2020, Tan2023, Ung2023} (see also \cite{TocchioPRL2014} for $1/2<n<2/3$). One should take into account that the common fixed-filling approach may require refinement since it does not naturally take into account phase-separation phenomenon where the specified total charge density exists within the phase-separated states only \cite{KapciaAPPA2016, KapciaAPPA2018}.

Note that the DMFT calculations can be cumbersome, time consuming, and with various convergence problems, especially with large number of self-consistency parameters. We thus limit the research scope of the present work: we focus of charge-ordering phenomenon neglecting possible magnetic orderings that are especially relevant when the onsite interaction dominates over the intersite interaction; the charge orders are commensurate with the $\sqrt{3}\times\sqrt{3}$ supercells; we consider the repulsive density-density interactions only; and the research was conducted for the ground state (zero temperature) only. 
We treat the intersite interaction on the mean-field (Hartree) level, hence spatial correlations are neglected in contrast to the local correlations.

Note, that in contrast to the local site-dependent DMFT \cite{AmaricciPRB2010,Kapcia2017}, there are several extensions containing the nonlocal intersite $V$ term in its construction as dynamical cluster approach \cite{MaierRMP2005,TerletskaPRB2018,PakiPRB2019,TerletskaPRB2021,IskakovPRB2022,KunduSP2024},
cellular DMFT \cite{KyungPRL2006,KyungPRB2006,MerinoPRL2007} as well as those combined with ab initio $GW$ scheme \cite{Ayralprl2012,Ayralprb2013,Huangprb2014}.
The other extensions of DMFT which are able to capture the nonlocal fluctuations are the dual-bosons approach \cite{LoonPRB2014} and the parquet approach \cite{PudleinerPRB2019,LI2019146}.
Most of these studies has been performed for EHM model on a 2D square lattice.
In the present work, however, we decided to neglect spatial correlations as we focused on the complex Mott physics occurring in the model.

The work is organized in the following way.
We describe the model  and the method of its solution (on a triangular lattice) in Sec. \ref{sec:method}. 
The results of the mean-field approximation \cite{AlekseevPRB2025} and the solution of the atomic-limit model \cite{Kapcia2021, KapciaJMMM2022} are found to be particularly useful in this research. We recap them in Sec. \ref{sec:recap} since we are extensively using them in the discussion of the results in Sec. \ref{sec:results}. We conclude with Sec. \ref{sec:summary} where we summarize the results and discuss the perspectives.

\section{Model and Method}\label{sec:method}

The extended Hubbard model with nearest-neighbor (NN) density-density intersite interaction (cf., e.g., \cite{MicnasRMP1990,LenacPRB1995,PietigPRL1999,TongPRB2004,AmaricciPRB2010,FerhatPRB2014,FevrierPRB2015,Kapcia2017,AlekseevPRB2025} has a form
\begin{equation}\label{eq:ham}\begin{aligned}
    \hat{H} 
    = &-t \sum_{\langle i,j \rangle, \sigma} \left( \hat{c}^\dag_{i\sigma} \hat{c}_{j\sigma} + \text{H.c.} \right)
    + U\sum_i \hat{n}_{i\uparrow}\hat{n}_{i\downarrow}
    \\&+ V \sum_{\langle i,j \rangle} \hat{n}_i \hat{n}_j
    - \mu\hat{N},
\end{aligned}\end{equation}
where 
$t$, $U$, $V$, and $\mu$ are a hopping amplitude, onsite and intersite density-density interaction strengths, and a chemical potential, respectively; 
$\hat{c}^\dag_{i\sigma}$ and $\hat{c}_{i\sigma}$ are creation and annihilation operators of a spin-$1/2$ fermion; 
$i=1,\ldots,L$ and $\sigma=\uparrow,\downarrow$ are lattice-site and spin indices, $L$ is a number of lattice sites enclosed by periodic boundary conditions; 
$\hat{n}_{i\sigma} = \hat{c}^\dag_{i\sigma} \hat{c}_{i\sigma}$ is an occupation-number operator, 
$\hat{n}_i = \sum_\sigma \hat{n}_{i\sigma}$, 
and $\hat{N} = \sum_i \hat{n}_i$. 
The summation $\sum_{\langle i,j \rangle}$ means the summation over all nearest-neighbor pairs (without repeating).

For the triangular lattice and with the assumption of periodicity of the $\sqrt{3}\times\sqrt{3}$ supercell (three sublattices with a sublattice index $\alpha=1,2,3$), the Fourier transform to the reciprocal space and the mean-field (Hartree) decoupling of the intersite interaction term (i.e., $\hat{n}_{i\sigma} \hat{n}_{j\sigma'} \overset{\text{MF}}{=} \langle\hat{n}_{i\sigma}\rangle \hat{n}_{j\sigma'} + \langle\hat{n}_{j\sigma'}\rangle \hat{n}_{i\sigma} - \langle\hat{n}_{i\sigma}\rangle \langle\hat{n}_{j\sigma'}\rangle$) turn the model (\ref{eq:ham}) to
\begin{equation}\label{eq:mf-ham}
\begin{aligned}
    \hat{H}
    = &\sum_{\mathbf{k}\sigma} \left(
    \varepsilon_\mathbf{k}
    (\hat{c}^\dag_{1\mathbf{k}\sigma}\hat{c}_{2\mathbf{k}\sigma}
    + \hat{c}^\dag_{3\mathbf{k}\sigma}\hat{c}_{1\mathbf{k}\sigma}
    + \hat{c}^\dag_{2\mathbf{k}\sigma}\hat{c}_{3\mathbf{k}\sigma})
    +\text{H.c.} \right)
    \\&
    - \sum_{\alpha\mathbf{k}\sigma} \mu_\alpha \hat{n}_{\alpha\mathbf{k}\sigma}
    + U\sum_i \hat{n}_{i\uparrow}\hat{n}_{i\downarrow}
    + C,
\end{aligned}\end{equation}
where 
$\hat{c}^\dag_{\alpha\mathbf{k}\sigma}$ and $\hat{c}_{\alpha\mathbf{k}\sigma}$ are creation and annihilation operators of a spin-$1/2$ fermion of a sublattice $\alpha$, momentum $\mathbf{k}$, and spin $\sigma$; the summation over $\mathbf{k}$ means the summation over the reduced first Brillouin zone (in total $L/3$ vectors); the Fourier-transform components
\begin{equation}\label{epsilon}
    \varepsilon_\mathbf{k} = - t \left(
    e^{i\mathbf{k}\mathbf{r}_1} + e^{i\mathbf{k}\mathbf{r}_2} + e^{i\mathbf{k}\mathbf{r}_3}
    \right)
\end{equation}
with
\begin{equation}
    \mathbf{r}_1 = \left(-\frac{1}{3},-\frac{2}{3}\right), \
    \mathbf{r}_2 = \left(-\frac{1}{3},\frac{1}{3}\right), \
    \mathbf{r}_3 = \left(\frac{2}{3},\frac{1}{3}\right),
\end{equation}
under the condition that the vectors $\mathbf{k}$ are written in the basis of the cell that is reciprocal to the $\sqrt{3}\times\sqrt{3}$ supercell; 
the constant $C = -\frac{L}{6} \sum_\alpha z V n_{\bar\alpha}n_{\bar{\bar\alpha}}$
and $\mu_\alpha = \mu - \frac{zV}{2}(n_{\bar\alpha} + n_{\bar{\bar\alpha}})$, where
$z=6$ is a coordination number, 
$\bar\alpha$ and $\bar{\bar\alpha}$ are sublattice indices different from $\alpha$ and from each other;
$\hat{n}_{\alpha\mathbf{k}\sigma}=\hat{c}^\dag_{\alpha\mathbf{k}\sigma}\hat{c}_{\alpha\mathbf{k}\sigma}$ and
$n_\alpha = \frac{3}{L}\sum_{\mathbf{k}\sigma} \langle \hat{n}_{\alpha\mathbf{k}\sigma} \rangle$. 
The total charge density is defined as
$n = \frac{1}{3} \sum_\alpha n_\alpha$.

Note, the mean-field decoupling of the intersite-interaction term is an exact approximation in the limit of the infinite coordination number which is the same limit where the DMFT is exact.

We express quantities in the half-bandwidth of the noninteracting triangular lattice $D = 4.5t$. Moreover, for better comparison between different lattices, we express intersite interaction parameter $V$ in units of $D/z = 0.75t$. While presenting results, we also make use of the shifted chemical potential $\bar\mu = \mu - U/2 - zV$. Note that there is no particle-hole symmetry around $\bar\mu = 0D$ due to the asymmetric noninteracting density of states of the triangular lattice, and it does not correspond to the half-filled ($n=1$) lattice \cite{AlekseevPRB2025}. Nevertheless, the form of the shifted chemical potential is chosen that way the $\bar\mu=0D$ is a point the particle-hole symmetry in the atomic limit (see sec. \ref{sec:recap}), and that way the mean-field contribution (aside from the constant term $C$) is eliminated for the non-charge-ordered lattice ($n_\alpha=n$) when $n=1$. More specifically, the constant shifted chemical potential keeps the non-charge-ordered solution with $n=1$ independent on $zV$ and $U$ in the mean-field approximation (particularly, at $\bar\mu_{n=1}\approx0.1855D$) \cite{AlekseevPRB2025} and independent on $zV$ within the DMFT ($\bar\mu_{n=1}\equiv\bar\mu_{n=1}(U)$ with $\bar\mu_{n=1}(0)\approx0.1855D$).

A matrix of Green's functions ($G_{\alpha\alpha'\mathbf{k}\sigma}$) of the model (\ref{eq:mf-ham}) is
\begin{equation}
    \mathbf{G}_{\mathbf{k}\sigma}(z) = 
    \begin{pmatrix}
        \xi_{1\sigma}(z) & -\varepsilon_\mathbf{k} & -\varepsilon_\mathbf{k}^* \\
        -\varepsilon_\mathbf{k}^* & \xi_{2\sigma}(z) & -\varepsilon_\mathbf{k} \\
        -\varepsilon_\mathbf{k} & -\varepsilon_\mathbf{k}^* & \xi_{3\sigma}(z)
    \end{pmatrix}^{-1},
\end{equation}
where
\begin{equation}
    \xi_{\alpha\sigma}(z) = z + \mu_\alpha - \Sigma_{\alpha\sigma}(z),
\end{equation}
and $\Sigma_{\alpha\sigma}(i\omega_n)$ is a sublattice-specific self-energy that within the DMFT is independent on $\mathbf{k}$.
Here, we consider the charge-ordering phenomena disregarding possible magnetic orders.
Hence, we have 
$G_{\alpha\alpha'\mathbf{k}\uparrow} = G_{\alpha\alpha'\mathbf{k}\downarrow} \equiv G_{\alpha\alpha'\mathbf{k}}$, 
$\Sigma_{\alpha\uparrow} = \Sigma_{\alpha\downarrow} \equiv \Sigma_\alpha$, 
and $\langle \hat{n}_{\alpha\mathbf{k}\uparrow} \rangle = \langle \hat{n}_{\alpha\mathbf{k}\downarrow} \rangle$.
A local Green's function is
\begin{equation}
    G_{\text{loc},\alpha}(z) = \frac{3}{L}\sum_\mathbf{k} G_{\alpha\alpha\mathbf{k}}(z),
\end{equation}
and DMFT self-consistency equations are
\begin{equation}
    \mathcal{G}_{0,\alpha}^{-1}(i\omega_n) = \Sigma_\alpha(i\omega_n) + G_{\text{loc},\alpha}^{-1}(i\omega_n),
\end{equation}
where $\omega_n = (2n+1)\pi/\beta$ is a Matsubara frequency, and the functions $\mathcal{G}_{0,\alpha}(i\omega_n)$ are noninteracting Anderson impurity model (AIM) Green's functions (or Weiss functions) for a sublattice $\alpha$, while the self-energies $\Sigma_\alpha(i\omega_n)$ are found from the AIMs of the previous self-consistency iteration. In our calculations we have $n_\text{s} = 9$ sites in the AIMs (i.e., $8$ bath sites) and use Lanczos algorithm to solve them ($800$ Lanczos iterations) \cite{GeorgesRMP1996} (cf. also \cite{CaffarelPRB1994,WeberPRB2012,AmaricciCPC2022,AmaricciSciPost2025}). For the Matsubara frequencies the parameter $\beta=1000/D$, while the maximal Matsubara frequency is $\omega_{8000} \approx 50.27D$. 

We use the grid of $96 \times 96$ vectors $\mathbf{k}$ ($817$ irreducible points), i.e., $L=288\times288$. Note that the triangular lattice with the symmetry, that is broken in accordance to the $\sqrt{3}\times\sqrt{3}$ supercell, has a wallpaper group $p3m1$ (number $14$), however, all $\mathbf{k}$-dependent quantities depend on two parameters
\begin{align}
    |\varepsilon_\mathbf{k}|^2 && \text{and} && \operatorname{Det}\left[
    \begin{pmatrix}
        0 & \varepsilon_\mathbf{k} & \varepsilon_\mathbf{k}^* \\
        \varepsilon_\mathbf{k}^* & 0 & \varepsilon_\mathbf{k} \\
        \varepsilon_\mathbf{k} & \varepsilon_\mathbf{k}^* & 0
    \end{pmatrix} \right]
\end{align}
which, luckily, have a symmetry of the symmetry-preserved triangular lattice ($p6mm$, number $17$). Hence, the irreducible number of vectors $\mathbf{k}$ in the reduced first Brillouin zone is smaller.

For the ground state (zero temperature) the grand potential of the model (\ref{eq:mf-ham}) is
\begin{equation}
\label{eq:omegaGS}
    \Omega = \langle\hat{H}\rangle = E_\text{kin} + E_\text{Hub} + E_\text{NN} + E_\text{chem},
\end{equation}
where the kinetic energy is calculated as
\begin{equation}\begin{aligned}
    E_\text{kin} = &\frac{4}{\beta}\operatorname{Re}\sum_{n\mathbf{k}}
    (
    \varepsilon_\mathbf{k} 
    [G_{21\mathbf{k}}(i\omega_n)
    + G_{13\mathbf{k}}(i\omega_n) 
    \\&+G_{32\mathbf{k}}(i\omega_n)]
    +\varepsilon_\mathbf{k}^*
    [G_{12\mathbf{k}}(i\omega_n) 
    \\&+ G_{31\mathbf{k}}(i\omega_n) + G_{23\mathbf{k}}(i\omega_n)]
    )  + E_\text{corr}
\end{aligned}\end{equation}
with $ E_\text{corr}$ as the correction to the kinetic energy due to the finite number of the Matsubara frequencies ($M=8000$),
i.e., due to the tail of the Green's functions (cf. \cite{GeorgesRMP1996,WeberPRB2012,AmaricciCPC2022}), is derived using the high-frequency behaviour of the self-energies ($\Sigma_\alpha(\infty) = \Sigma_\text{mean-field} = \frac{U}{2}n_\alpha$) and for the triangular lattice can be expressed as
\begin{equation}
    E_\text{corr}
    =-L\frac{6\beta t^2}{\pi^2} \psi'\left(\frac{3}{4}+M\right)
    \approx -L\frac{6\beta t^2}{M\pi^2},
\end{equation}
where $\psi'(z)$ is a polygamma function of the first order.
The next terms in (\ref{eq:omegaGS}) are:
the onsite interaction (Hubbard) energy
\begin{equation}
    E_\text{Hub} = L\frac{U}{3}\sum_\alpha \left<\hat{n}_{\alpha\uparrow}\hat{n}_{\alpha\downarrow}\right>
\end{equation}
with the double occupancy $\left<\hat{n}_{\alpha\uparrow}\hat{n}_{\alpha\downarrow}\right>$ found from the AIMs; the nearest-neighbor intersite interaction energy
\begin{equation}
    E_\text{NN} = L\frac{zV}{6} \sum_\alpha n_{\bar\alpha}n_{\bar{\bar\alpha}};
\end{equation}
and the chemical energy
\begin{equation}
    E_\text{chem} = -L \mu n.
\end{equation}

There are severe problems in convergence of the DMFT self-consistency algorithm across the whole phase diagram, excluding insulating phases. 
Some convergence problems are similar to those in the mean-field approximation (MFA) \cite{AlekseevPRB2025}, where the Fermi level is located near the peak or singularity of the spectral function, but in the MFA the locations of these peaks are well-defined and easily understood from band structures. 
Mixing procedures for the sublattice occupation numbers, sublattice self-energies (or $\xi_{\alpha\sigma}(i\omega_n)$), and Weiss functions (or AIM hybridization functions) are vastly varied across the phase diagram. 
For some points of the phase diagram we were not able to rich convergence at all, however, the algorithm can often be stabilized by including the thermally excited states while still staying close to the ground state (\mbox{$T=\beta^{-1}=10^{-3}D$}). 
Moreover, for the same phase and same point on the phase diagram, there can be found slightly different self-consistent solutions. 
Thus, the points of phase transitions could also slightly vary, while the difference in grand potentials is too small to draw conclusions on what phases are stable or metastable (even the summation over a finite number of points $\mathbf{k}$ implies inability to compare these small differences in $\Omega/L$, as it was found for the Bethe lattice in the MFA \cite{AlekseevPhysicaA2026}). 
These problems slow down the calculations, but make difference on a small scale only, meanwhile, we are confident in the found general picture and the shown phase diagram.

Both AIM solver (Lanczos algorithm) and DMFT algorithm used in this work are implemented in our in-house codes in the \texttt{python} programming language.

\section{Atomic-Limit and Mean-Field Results}\label{sec:recap}

\begin{figure}
    \centering
    \includegraphics[width=1\linewidth]{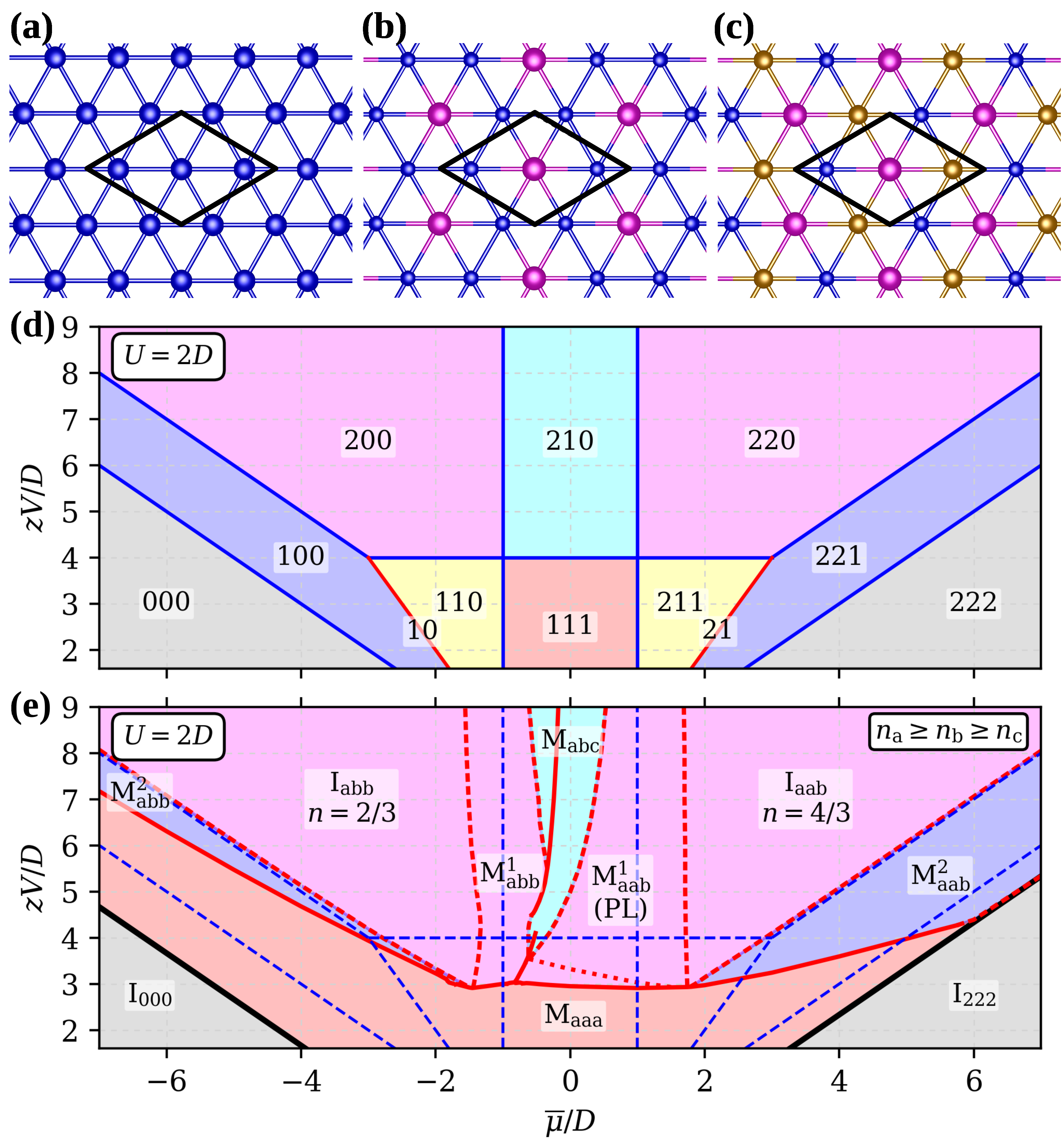}
    \caption{(a)--(c) Charge order types of $\sqrt{3}\times\sqrt{3}$ supercell, where the supercell is shown with a black solid line and the lattice sites with different color and size have different occupation numbers. 
    (d) Atomic-limit phase diagram, where blue and red solid lines are phase transitions, the latter are also lines of stability of the stripe order (cf. \cite{Kapcia2021}). 
    (e) Mean-field approximation phase diagram, where solid and dashed red lines are for discontinuous and continuous phase transitions, respectively (cf. \cite{AlekseevPRB2025}). 
    Below a red dotted line (approximately), the M$^1_\text{aab}$ phase is not a PL phase anymore.
    The AL lines are also shown on the MFA phase diagram with dashed blue lines.
    Colors of regions and phase notations are adapted in the light of the following DMFT results.
    }
    \label{fig:recap}
\end{figure}

In this section we give a short recap of atomic-limit (AL) \cite{Kapcia2021,KapciaJMMM2022} and mean-field approximation \cite{AlekseevPRB2025} ground-state results for a triangular lattice which are extensively referred throughout the paper.

The atomic-limit calculations is a simple and useful tool that gives the $t \rightarrow 0$ or $U \gg t$, $zV \gg t$ limit. The phases in the AL can be named by three integer occupation numbers of three sublattices. Note that as far as a periodic supercell is chosen, the solution of the AL model (eq. (\ref{eq:ham}) with $t=0$) is exact: not even spacial correlations and magnetic orderings are neglected, in contrast to our following DMFT results. However, no mechanism of spontaneous symmetry breaking is inherently introduced, and, for example, the phase $110$ includes phases like $(\uparrow,\uparrow,0)$, $(\uparrow,\downarrow,0)$, and its total degeneracy is $12$. The AL phase diagram \cite{Kapcia2021} for $U=2D$ is shown in Fig. \ref{fig:recap}d. Here, $D$ are arbitrary units (since defined for the rest of the paper $D = 4.5t$ equals zero in the AL) in which the onsite interaction strength $U$ is initially expressed (here, $U=2D$). Such an approach gives a noticeably good outline of the further $\bar\mu$-$V$ phase diagrams where $D$ is a half-bandwidth. The $U=0D$ is the limiting case with only $4$ phases in the AL phase diagram: $000$, $200$, $220$, and $222$.

In the MFA, both $U$ and $V$ terms of the model (\ref{eq:ham}) are considered on the Hartree level, and consequently, it cannot yield Mott physics. Meanwhile, the analysis of band structures available from the MFA provides clarity about a number of phase-diagram features found within the DMFT; it allows to identify what features come from strong correlations; and overall useful considering small computational and time requirements, despite convergence problems that appear already on the mean-field level \cite{AlekseevPRB2025}. In this work, the intersite interaction is also considered on the Hartree level. Hence, the $U \rightarrow 0$ limit of our model has been found in the MFA (see Fig. 2 in Ref. \cite{AlekseevPRB2025}). This work can be viewed as an extension of the work in Ref. \cite{AlekseevPRB2025}. 

Fig. \ref{fig:recap}e shows the MFA phase diagram for $U=2D$ \cite{AlekseevPRB2025}. The most basic step towards phase identification is the type of charge order (or its absence), as shown in Figs. \ref{fig:recap}a--\ref{fig:recap}c. The occupation numbers of three sublattices ($n_1$, $n_2$, $n_3$) of the $\sqrt{3}\times\sqrt{3}$ supercell can all take the same values preserving the initial triangular-lattice symmetry ($n_1=n_2=n_3$, Fig. \ref{fig:recap}a), can take two different values ($n_1=n_2 \ne n_3$, Fig. \ref{fig:recap}b), 
and can all take different values leading to the most broken symmetry configuration ($n_1 \ne n_2 \ne n_3 \ne n_1$, Fig. \ref{fig:recap}c). The phases can also be identified as metals (M) and insulators (I), while a metallic phase can also be a pinball (PL) liquid phase. The subscript in the phase names on the MFA phase diagram (Fig. \ref{fig:recap}e) denotes the three occupation numbers, e.g., an aab phase means $n_1=n_2=n_\text{a}$ and $n_3=n_\text{b}$. While both aab and abb phases belong to the type of charge order depicted in Fig. \ref{fig:recap}b, they are quite different phases (as follows from band structure analysis) with a discontinuous transition between each other. The way to distinguish them is that $n_\text{a}>n_\text{b}$, specifically, the aab phases have two charge-rich and one charge-poor sublattice, while the abb phases have one charge-rich and two charge-poor sublattices.

The AL model (eq. (\ref{eq:ham}) with $t=0$) has a clear particle-hole symmetry, i.e., a symmetry of the phase diagram with respect to $\bar\mu=0D$. Meanwhile, due to an asymmetric noninteracting density of states of the triangular lattice (which in turn comes from geometrical frustration or nonbipartiteness \cite{hanisch1997}), the MFA phase diagram has a strong particle-hole asymmetry manifested in nearly every phase and phase transition \cite{AlekseevPRB2025}.
We show the correspondence between AL and MFA phases with colors that are also used for the following DMFT results: $111$ region (red), $110$ and $211$ regions (yellow), $100$ and $221$ regions (blue), $200$ and $220$ regions (magenta), $210$ region (cyan), and $000$ and $222$ regions (gray). The latter are fully unoccupied and fully occupied phases with well-determined borders (AL: $|\bar\mu| = \frac{U}{2} + zV$; MFA: $\bar\mu = -(zV + \frac{U}{2} + 6t)$ and $\bar\mu = zV + \frac{U}{2} + 3t$).
In the large-$zV$ limit, the correspondence between the AL and MFA results is prominent. As follows from the AL results, the red, cyan, blue, and yellow regions ($111$, $210$, $100$, $221$, $110$, and $211$ regions, i.e. those that have $1$ in the name) come from nonzero $U$, and hence, likely require consideration of the Mott physics to correctly describe them. Note, the yellow regions both require Mott physics and do not exist in the large-$zV$ limit ($zV \le 2U$ for the AL), and as a result, they do not appear on the MFA phase diagram at all.

In the AL, the charge order exists for any nonzero $zV$, while in the MFA is starts from $zV \approx U + D$ only. In contrast to a Bethe lattice \cite{Kapcia2017,AlekseevPhysicaA2026}, the symmetry breaking from the non-charge-ordered metal to the charge-ordered metal is a discontinuous transition in the MFA, except for a few small regions of the chemical potential $\bar\mu$ which has been explored by means of the band-structure analysis~\cite{AlekseevPRB2025}. 

We also show regions of stability of a stripe order \cite{mahmoudian2015glassy, jin2021stripe, Tan2023} on the AL phase diagram, denoted with $10$ and $21$. 
The stripe order is not commensurate with the supercell $\sqrt{3}\times\sqrt{3}$, and it would require a supercell with $6$ lattice sites to capture both stripe order and orders in Figs. \ref{fig:recap}b and \ref{fig:recap}c, heavily compromising both convergence stability and computational time in the DMFT. 
In the AL, stripes are stable in the line of transition between the $100$ and $110$ phases or the $221$ and $211$ phases, thus the three phases are degenerate on these lines (have the same grand potential). 
Note that the stripes were not Ref.~in \cite{Kapcia2021}.

The PL phase is found in the MFA on one side of the phase diagram (an aab phase), because on the other side of the phase diagram in the corresponding phase, the Fermi level is located in the same energy region where the hybridization between sublattices (honeycomb and triangular sublattices that are shown in Fig. \ref{fig:recap}b) is concentrated \cite{AlekseevPRB2025}.

More references to the results of the MFA calculations are used throughout the paper in the following parts.

\section{Phase Diagram}\label{sec:results}

\begin{figure*}
    \centering
    \includegraphics[width=1\linewidth]{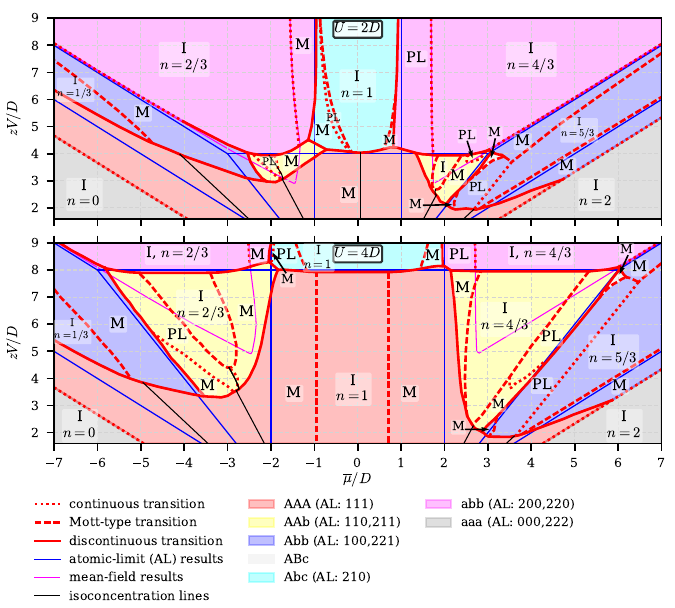}
    \caption{DMFT phase diagram of the triangular-lattice extended Hubbard model with charge orders commensurate with $\sqrt{3}\times\sqrt{3}$ supercell. Only the phases with lowest grand potential are shown (no coexistence regions).
    On the phase diagram, M, I and PL are metallic, insulating and pinball-liquid phases, respectively; the isoconcentration lines from left to right correspond to $n=1/3$, $2/3$, $1$, $4/3$, and $5/3$. 
    In notations of colored regions, the small letters (a, b, c) are for weakly correlated sublattices ($Z_\alpha \approx 1$ and sublattices can be described in the MFA), and the capital letters (A, B) are for strongly correlated sublattices with Mott physics. 
    The mean-field results for the phases I$_\text{abb}$ and I$_\text{aab}$ (Fig. \ref{fig:recap}e), and the atomic-limit results (Fig. \ref{fig:recap}d) are also shown for the sake of analysis.}
    \label{fig:phase-diagram}
\end{figure*}

The DMFT phase diagram of the triangular-lattice EHM with charge orders that are commensurate with the $\sqrt{3}\times\sqrt{3}$ supercell is shown in Fig. \ref{fig:phase-diagram}, particularly, the two-dimensional cross-sections for $U=2D$ and $4D$ of the three-dimensional phase diagram (this is the central result of the work). 
We discuss the phase naming and order parameters in Sec. \ref{sec:phases} and the found phase transitions in Sec. \ref{sec:transitions}. 
Next, we pay special attention to the mean-field features of the phase diagram (Sec. \ref{sec:MFA}) and pinball-liquid phases (Sec. \ref{sec:PL}).

\subsection{Phase Identification}\label{sec:phases}

\begin{figure}
    \centering
    \includegraphics[width=1\linewidth]{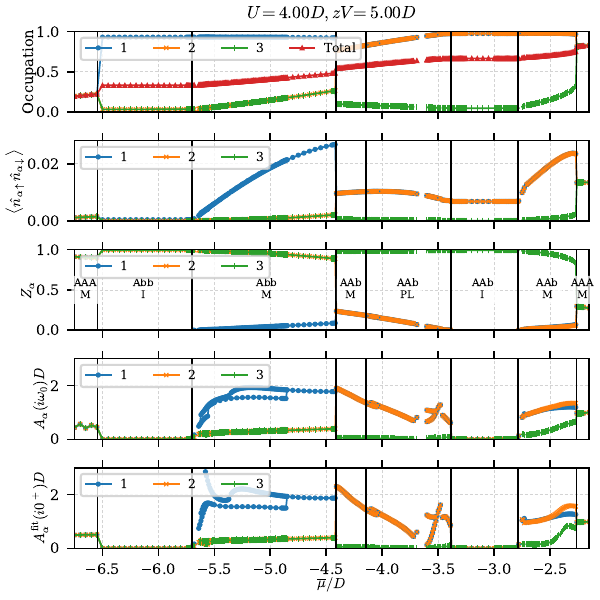}
    \caption{
    Order parameters along the line $U=4D$ and $zV=5D$ encompassing the AAb-region charge-transfer-driven pinball liquid and three Mott-type transitions ($\bar\mu=-5.70D$, $-3.385D$, and $-2.785D$). 
    Legend shows the index of sublattice $\alpha$. 
    Coexistence regions are not shown (only phases with lowest grand potential). 
    Points where the lines are interrupted are where convergence has not been reached. 
    Vertical solid lines show phase transitions. 
    Both spectral weights $A_\alpha$ at the Matsubara frequency $i\omega_0$ and the one taken from fitting with rational function at $i0^+$ are shown: the one from the fitting is generally more realistic but the fitting is unstable in some cases (see Fig. \ref{fig:M_to_MottPL}).
    Different solutions at ranges of $\bar\mu \approx -3.6D$, $-4.1D$, $-5.6D$, and a large region form $-5.6$ to $-4.9D$ are discussed at the end of Sec. \ref{sec:transitions}.
    }
    \label{fig:Abb_to_AAb}
\end{figure}

\begin{figure}
    \centering
    \includegraphics[width=1\linewidth]{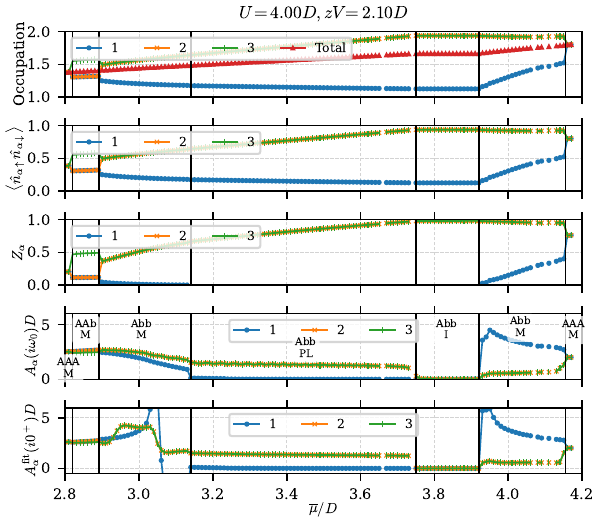}
    \caption{
    Order parameters along the line $U=4D$ and $zV=2.1D$ encompassing the Abb-region Mott-localization-driven pinball liquid. 
    The fitting for $A_\alpha$ is unstable around the transition from the metallic to pinball-liquid Abb phase.
    $Z_\text{b}$ (i.e., $Z_2$ and $Z_3$ for Abb region and $Z_3$ for AAb region) is not close to $1$ due to the proximity to the point of the continuous transition to the AAA region.
    See the caption of Fig. \ref{fig:Abb_to_AAb} for more details. 
    }
    \label{fig:M_to_MottPL}
\end{figure}

Same as in the MFA (Sec. \ref{sec:recap}), the basic step in phase description is to identify its charge-order type (Figs. \ref{fig:recap}a--\ref{fig:recap}c; order parameters $n_1$, $n_2$, and $n_3$). The phase diagram (Fig. \ref{fig:phase-diagram}) is divided into color regions where regions aaa (gray) and AAA (red) contain non-charge-ordered phases (Fig. \ref{fig:recap}a), regions AAb (yellow), Abb (blue), and abb (magenta) contain phases with charge order from Fig. \ref{fig:recap}b, and regions ABc (white) and Abc (cyan) are for phases with the most broken symmetry (Fig. \ref{fig:recap}c). 

The type of the charge order cannot fully describe the region on its own. Another criterion to distinguish the found phases is whether one, two, or all three sublattices experience strong onsite correlations, (i.e., the Mott physics), or otherwise, could have been well-described in the MFA being weakly correlated. To quantify this criterion we use three more order parameters, $Z_1$, $Z_2$, and $Z_3$, the sublattice-specific renormalization constants, also known as quasiparticle weights:
\begin{equation}
    Z_\alpha = \left( 1 - \left.\frac{d\operatorname{Im}\Sigma_\alpha(i\omega)}{d(i\omega)}\right|_{\omega=0} \right)^{-1}.
\end{equation}
Particularly, $Z_\alpha \approx 1$ for weakly correlated sublattices and $0 \le Z_\alpha < 1$ for strongly correlated sublattices.
We use capital letters (A, B) to show that a sublattice is strongly correlated and small letters (a, b, c) otherwise. Thus, for example, the region AAb contains phases characterized by two sublattices that have the same occupation numbers ($n_1=n_2=n_\text{A}$) and exhibit Mott physics (i.e., strongly correlated), and one sublattice with another occupation number ($n_3=n_\text{b}$) that could have been described in the MFA (i.e., weakly correlated). 
In practice, we find $Z_\alpha$ from a quadratic approximation of $\Sigma_\alpha$ with $i\omega_0$ and $i\omega_1$ assuming $\Sigma_\alpha(i0)=0$.
The renormalization constants $Z_\alpha$ are not calculated for sublattices where the Mott localization takes place, however, no ambiguity appears in the process of decision whether such a sublattice is strongly correlated. In such a case, the continuous metal-insulator or metal-pinball-liquid transition with $Z_\alpha \rightarrow 0$ is present on the phase diagram.
Moreover, three more order parameters, double occupancies, 
$\langle \hat{n}_{1\uparrow}\hat{n}_{1\downarrow} \rangle$, $\langle \hat{n}_{2\uparrow}\hat{n}_{2\downarrow} \rangle$, and $\langle \hat{n}_{3\uparrow}\hat{n}_{3\downarrow} \rangle$ 
that are found from solution of the AIMs, help with identification whether a sublattice is strongly correlated. Particularly, when the Mott localization takes place on a sublattice $\alpha$, its double occupancy ($\langle \hat{n}_{\alpha\uparrow}\hat{n}_{\alpha\downarrow} \rangle$) is close to $0$ despite the fact that its occupancy ($\langle\hat{n}_\alpha\rangle$) is close to $1$.

When analyzing the found and described above regions, the comparison with the AL phase diagram proves useful (blue dashed lines in Fig. \ref{fig:phase-diagram}; see also Fig. \ref{fig:recap}d and its colored phases). Except for a small unexpected ABc region and a nearly filled metallic Abb phase, all regions are nicely outlined by the AL results, where sublattices with the occupation number $1$ in the AL appear to be strongly correlated within the DMFT (represented by capital letters A and B), in contrast to those with occupation numbers $0$ and $2$ (represented by small letters a, b, and c; see legend in Fig. \ref{fig:phase-diagram}).

Finally, having identified the regions, each of them can have phases that are insulating (I), metallic (M), or pinball liquid (PL). Hence, we introduce the finial three order parameters, sublattice-projected spectral weights at the Fermi level, 
\begin{equation}
    A_\alpha(i0^+) = \lim_{\eta\rightarrow+0D} A_\alpha(i\eta),
\end{equation}
where
\begin{equation}
    A_\alpha(z) = -\frac{2}{\pi} \operatorname{Im} G_{\text{loc},\alpha}(z)
\end{equation}
is a sublattice-projected spectral function. The metallic phases are characterized by $A_\alpha(i0^+)D > 0$ for all sublattices. They appear in every region of the phase diagram except for the aaa (gray) region, i.e., the region of fully unoccupied and fully occupied lattices. The insulating phases have $A_\alpha(i0^+)D = 0$ for all sublattices and strictly integer value of $3n$ (i.e., $\sum_\alpha n_\alpha$). Even though it is numerically unstable to take the limit $\eta \rightarrow +0$ for $A_\alpha(i\eta)$, the location of the metal-insulator transition is always clear from the results of the DMFT calculations. Insulating phases (both charge-ordered and non-charge-ordered) can be band insulators, where all three sublattices are weakly correlated or even not correlated at all (see Sec. \ref{sec:MFA}), and can be insulators with Mott localization, as found from the behaviour of $Z_\alpha$ at the metal-insulator transition and from $\langle \hat{n}_{\alpha\uparrow}\hat{n}_{\alpha\downarrow} \rangle$, as discussed above (see, for example, three metal-insulator transitions in Fig. \ref{fig:Abb_to_AAb}).

Finally, the pinball-liquid phases are characterized by $A_\alpha(i0^+)D > 0$ for two sublattices and $A_\alpha(i0^+)D = 0$ for the third one (see Sec. \ref{sec:PL} for their discussion). 
We distinguish two kinds of PL phases: one where the sublattice with ``pins'' is insulating due to the Mott localization, and another that we can refer as a charge-transfer-driven PL, where the localization on ``pins'' is due to domination of the intersite repulsion over the itineracy.
When $A_\alpha(i0^+)D = 0$ for one particular sublattice $\alpha$ is because it experiences the Mott localization, the transition between a metal and a pinball liquid is found by following the behaviour of $Z_\alpha$ along this transition (Fig. \ref{fig:M_to_MottPL}, $\bar\mu=3.14D$), and the transition involves the coexistence region (hysteresis). 
Meanwhile, the identification of the charge-transfer-driven PL requires more careful considerations. 
First, using the obtained during DMFT self-consistency algorithm $\Sigma_\alpha(i\omega_n)$, we recalculate $G_{\text{loc},\alpha}(i\omega_n)$ using a denser $\mathbf{k}$-point grid of $480 \times 480$. 
It is done to correct possible numerical inaccuracies at very small but important Matsubara frequencies. 
Next, we fit the whole function $G_{\text{loc},\alpha}(i\omega_n)$ (both real and imaginary part) from $\omega_0$ to $\omega_{8000}$ with a rational function of the order $4/4$, i.e., 
$\left( \sum_{i=0}^4 a_i \omega^i \right) / \left( 1 + \sum_{i=1}^4 b_i \omega^i \right)$. 
It is found to give a more numerically stable result than fitting only the first few Matsubara frequencies with polynomial functions. 
Finally, we evaluate $G^\text{fit}_{\text{loc},\alpha}(i0^+) = a_0$ and consider a phase to be a pinball liquid if for one of the sublattices $\operatorname{Im}G^\text{fit}_{\text{loc},\alpha}(i0^+)D > -0.05$ or equivalently $A^\text{fit}_\alpha(i0^+)D < 0.032$.

To conclude this section, we should also mention that besides its value on the Fermi level, we also use the imaginary part of the spectral function on the whole real axis ($\operatorname{Im}A_\alpha(\omega+0.03i)$) to analyze and describe the found phases (see the Supplemental Material (SM)~\footnote{See Supplemental Material at [URL will be inserted by publisher] for a few representative sublattice-projected spectral functions of the phases found in the model studied.}). 
Particularly, see Fig. \ref{fig:AAb} for the AAb-region phases, Fig. \ref{fig:Abb} for the Abb-region phases, Fig. \ref{fig:Abc} for the Abc-region phases, and Fig. \ref{fig:abb} for the abb-region phases as well as their discussion in the SM \cite{Note1}.

\subsection{Phase Transitions and Quarter Filling}\label{sec:transitions}

\begin{figure}
    \centering
    \includegraphics[width=1\linewidth]{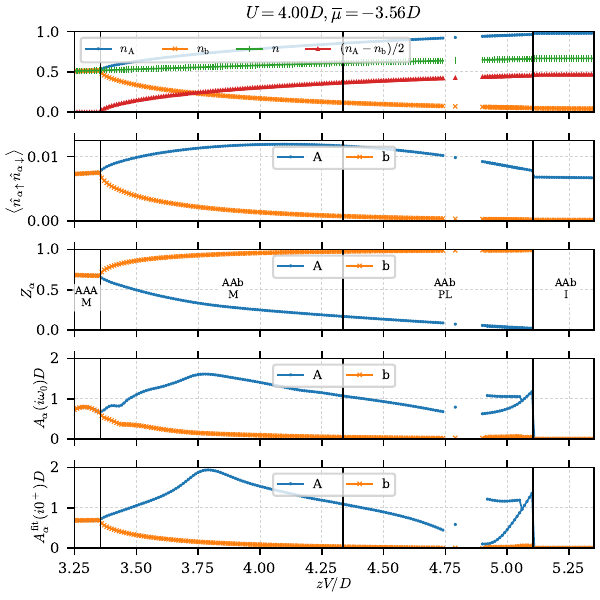}
    \caption{
    Order parameters for the AAb phases along the line $U=4D$ and $\bar\mu=-3.56D$ encompassing the point of continuous symmetry breaking at $zV=3.355D$ (compare with discontinuous symmetry breakings on the edges of Figs. \ref{fig:Abb_to_AAb} and \ref{fig:M_to_MottPL}), the charge-transfer-driven pinball liquid, as well as the Mott-type transition at $zV=5.105D$. 
    Different metallic solutions at a range of $zV$ from $4.92D$ to $5.05D$ are discussed at the end of Sec. \ref{sec:transitions}.
    See the caption of Fig. \ref{fig:Abb_to_AAb} for more details.
    }
    \label{fig:continuous_sym_break}
\end{figure}

\begin{figure}
    \centering
    \includegraphics[width=1\linewidth]{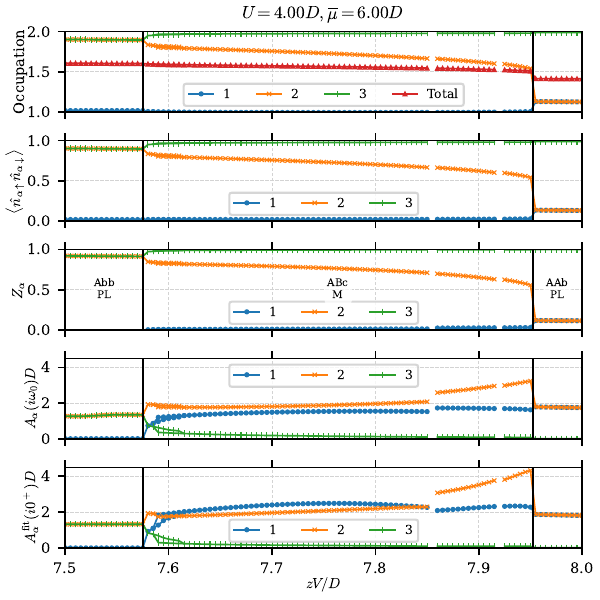}
    \caption{
    Order parameters along the line $U=4D$ and $\bar\mu=6D$ encompassing the ABc region (intermediate region between AAb and Abb regions) and two pinball-liquid phases. 
    Different solutions at a range of $\bar\mu \approx 7.6D$ are discussed at the end of Sec. \ref{sec:transitions}.
    See the caption of Fig. \ref{fig:Abb_to_AAb} for more details.
    }
    \label{fig:ABc}
\end{figure}

\begin{figure}
    \centering
    \includegraphics[width=1\linewidth]{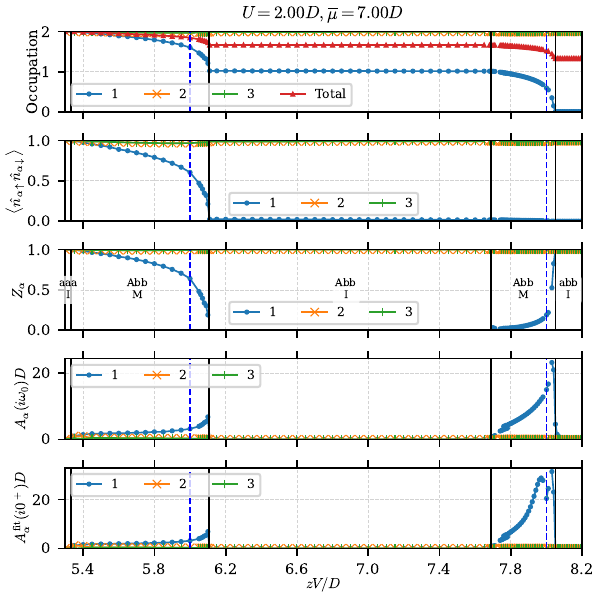}
    \caption{
    Order parameters along the line $U=2D$ and $\bar\mu=7D$ encompassing the continuous transitions of the Abb-region metals to the fully occupied phase and the abb-region band insulator. 
    The $A_\text{b}(i0^+)D$ (i.e., $A_2(i0^+)D$ and $A_3(i0^+)D$) in both Abb-region metals is nonzero ($\sim 0.1$--$0.5$) which is unclear from the shown scale; in fact, the itineracy between sites of the sublattice A can only take place through the hopping with the honeycomb sublattice b. Blue dashed vertical lines are for the atomic-limit $222$-$221$ and $221$-$220$ phase transitions. 
    Different solutions at a range of $zV \approx 7.77D$ are discussed at the end of Sec. \ref{sec:transitions}.
    See the caption of Fig. \ref{fig:Abb_to_AAb} for more details.
    }
    \label{fig:aaa_Abb_abb}
\end{figure}

To make the phase diagram more informative, we distinguish three types of phase transitions: discontinuous (first order), continuous (second order), and Mott-type transitions. 
The Mott-type transitions are characterized by the approach of one or more of sublattice-specific quasiparticle weights $Z_\alpha$ to zero on a metallic side of the transition (see various Mott-type transitions in Figs. \ref{fig:Abb_to_AAb}--\ref{fig:aaa_Abb_abb}; cf. also discussion of the transition in the absence of long-range order in \cite{GeorgesRMP1996,ImadaRMP1998}).
Note that without the charge order and on a Bethe lattice, the metal-to-Mott-insulator transition is discontinuous for finite temperatures, but becomes effectively continuous in the ground state (the scope of this research) while still having a coexistence region where the Mott-insulating phase is metastable \cite{GeorgesRMP1996}. 

Noticeably, the Mott-type transitions in the Abc (cyan) region and the Mott-type transitions of the Abb (blue) regions between the insulator and the metal with the larger $zV$ (not the one that is near the fully occupied phase) are purely continuous without any hysteresis (coexistence) between the phases. The rest of the Mott-type transitions shown in Fig. \ref{fig:phase-diagram} are more typical Mott transitions with hystereses, however, the point of transition (i.e., the point where grand potentials of two phases are equal) is not necessarily at the edge of the coexistence region (in contrast to the non-charge-ordered Bethe lattice \cite{GeorgesRMP1996}). For example, one can clearly see a jump of $\langle\hat{n}_{\text{A}\uparrow}\hat{n}_{\text{A}\downarrow}\rangle$ in Fig. \ref{fig:continuous_sym_break} (coexistence regions are not shown) at $zV=5.105D$: the transition from the metallic phase can go continuously but we mark the transition before it happens.
It may both be a real phenomenon or come from numerical problems when comparing small differences between grand potentials, due to the sum over limited number of $\mathbf{k}$ points. As found in Ref. \cite{AlekseevPhysicaA2026}, these numerical problems are hard to overcome even for simpler lattices in the MFA, unless analytical simplifications are used.

The discontinuous transitions are the most common transitions between the colored regions with the exception of the aaa (gray) regions. 
The discontinuous transitions imply the existence of jumps in the total charge density $n$, and hence, the existence of phase-separated states that host an intermediate charge density $n$ and are poorly captured within the common fixed-density approaches.
Worth noting that the symmetry breaking from the non-charge-ordered metal phase of the AAA region to the charge-ordered metal phases of AAb and Abb regions can happen continuously in small parameter ranges, as discussed in Sec. \ref{sec:MFA} and shown in Fig. \ref{fig:continuous_sym_break}. 

We found a new metal ABc phase (the only phase in the ABc region) with the most broken symmetry (Fig. \ref{fig:recap}c). It can be viewed as an intermediate phase between AAb and Abb regions, having the discontinuous transitions with the AAb and Abb metallic phases and a Mott-type transition with the Abb pinball-liquid phase, see Fig. \ref{fig:ABc}. Extent of the discontinuity (how large the jump of the order parameters) of the transition to the AAb region also varies, potentially becoming a continuous transition for a small range of parameters, particularly, it is rather continuous along the line with $U=2D$ and $zV=4D$, where $4$ phases are close to each other (ABc metal, AAb metal, Abb metal, and abb insulator).
Note that its transition to the Abb pinball-liquid phase is depicted rather approximately due to the complication of its analysis that comes from the existence of various solutions and convergence problems.

A quarter filling, i.e., $n=1/2$ for electron quarter filling and $n=3/2$ for hole quarter filling, is worth special attention since it is the filling found in triangular-lattice organic conductors \cite{mori1998systematic, seo2004toward} (cf. also \cite{Merino2013}). 
Note that $n=1/2$ and $n=3/2$ are exchangeable with $t \rightarrow -t$ in the Hamiltonian (\ref{eq:ham}). Note also that in the MFA (and hence for $U=0D$), the charge density $n=3/2$ corresponds to the location of the Fermi level at the Van Hove singularity (for a non-charge-ordered phase). 
In the DMFT, the quarter filling is located close to the AL lines between the 100 and 110 (221 and 211) phases and between the 100 and 200 (221 and 220) phases. The most of the quarter filling takes place in the Abb region, both in the metal (for $n=1/2$ and $3/2$) and pinball-liquid ($n=3/2$ only) phases. 
However, for a range of parameters, the quarter filling can exist only within the phase-separated states between the AAb and Abb regions and within the ABc region. Moreover, the AL results make us expect the possible stability of the stripe order approximately along the line of the quarter filling (cf. Fig. \ref{fig:recap}d) implying even more complex behaviour for the organic conductors. 
For clarity and consistence of the present work, a more detailed analysis of the quarter filling is deferred to a future paper.

We should also mention that the phase diagram (within each metallic and pinball-liquid phase) is also full of small discontinuous (with coexistence regions) transitions between essentially the same phase (see such regions in Figs. \ref{fig:Abb_to_AAb}, \ref{fig:continuous_sym_break}--\ref{fig:aaa_Abb_abb}; not shown in Fig. \ref{fig:phase-diagram}). 
The main difference between two such solutions is in magnitudes of $A_\alpha(i0^+)$, and the character of these transitions is very similar to the transition found in the MFA between the abb-like M$_\text{abc}$ phase and aab-like M$_\text{abc}$ phase (see Fig. \ref{fig:recap}e, inside the cyan region). 
In the MFA results, the Fermi level goes through a peak in the spectral function and the phase is slightly different on different sides of the peak with hysteresis between these solutions, see Ref. \cite{AlekseevPRB2025}. 
The grand-potential difference within hysteresis between such solutions is also too small to conclude which of them is more stable. 
These ``phase transitions'' can also have a critical point where the transition and the difference between the two solutions disappear (once again, similar to the MFA results---see the end of the phase transition within the cyan region at around $zV=4D$ in Fig. \ref{fig:recap}e).

\subsection{Mean-Field Features and Consequences of the Mott Physics}\label{sec:MFA}

A number of features are easier to interpret in the MFA while they still appear within the DMFT. This section is devoted to discuss these features.

Most importantly, we found that the weakly correlated sublattices (those denoted by small letters a, b, and c) are perfectly reproduced in the MFA if the phase is insulating, i.e., for insulating phases such sublattices are noncorrelated at all. Thus, the insulators of the abb and aaa regions are band insulators and can be investigated without the DMFT, since we still treat the intersite interaction on the mean-field Hartree level. For this reason, the borders of the aaa-region and abb-region insulators are the same as in the MFA \cite{AlekseevPRB2025}; their knowledge help to preform cumbersome DMFT research; and these phases can be analyzed by following their mean-field band structures \cite{AlekseevPRB2025}. 

As easily found in the MFA, being a noncorrelated insulating phase of the aaa region, the fully unoccupied phase has a continuous transition to the non-charge-ordered metal phase (AAA) at $\bar\mu = -\left(zV + \frac{U}{2} + 6t\right)$. The same way, the fully occupied phase has a continuous transition to both AAA and Abb regions at $\bar\mu = zV + \frac{U}{2} + 3t$. 
In general, the strong particle-hole asymmetry found in the MFA due to the asymmetrical noninteracting density of states is even stronger within the DMFT. Moreover, the electron-doped ($\bar\mu>0D$) side of the phase diagram is more affected by strong correlations, particularly, the charge-ordered phases appear for wider range of $zV$, and it hosts a PL phase with Mott localization (see Sec. \ref{sec:PL}). Note that the electron-doped side is the side where the Van Hove singularity is located in the non-charge-ordered mean-field (or $U=0D$) solutions.

The nontypical continuous transition from the fully occupied phase directly to the charge-ordered phase (Fig. \ref{fig:aaa_Abb_abb}, $zV=5.33D$) has been discussed in more details in the MFA \cite{AlekseevPRB2025} and is the same within the DMFT. The difference is that it starts at a lower $zV$, just like the charge-ordered phases in general---in the MFA they appear only for $zV \gtrsim U + D$. This nearly filled metal phase of the Abb region is worth special attention. It appears already for $U=0D$ \cite{AlekseevPRB2025} even though there is no counterpart of this phase in the AL for $U=0D$ (no $221$ phase). Thus, this phase is purely a result of finite itineracy and is a feature of the frustrated triangular lattice that does not depend on strength of the onsite correlations. In the MFA, the phase can become a narrow-band metal (M$^2_\text{aab}$ in Fig. \ref{fig:recap}e) where the Fermi level is located in an atomic-like band (nearly a level) of the sublattice b that is well-separated from honeycomb-like bands of the sublattice a. The metallic properties come from very weak hybridization between the sublattices a and b---an effective hopping between the sublattice b through the honeycomb sublattice. Naturally, within the DMFT and $U>0D$, this atomic-like level gets separated into two Hubbard bands (the insulating phase of the Abb region, Fig. \ref{fig:aaa_Abb_abb}, $zV=6.105D$). The transition to the Abb insulating phase takes place almost immediately after the line of $222$-$221$ transition of the AL phase diagram, thus, the nearly filled charge-ordered metallic phase is located between well-defined lines of the MFA and AL transitions to the fully occupied phase: $\bar\mu = zV + \frac{U}{2} + 3t$ and $\bar\mu = zV + \frac{U}{2}$. The renormalization constant $Z_\text{A}$ changes from $1$ at the continuous transition to the aaa range, to $0$ at the discontinuous Mott-type transition to the Abb insulator, and the sharpest change of $Z_\text{A}$ is around the AL $222$-$221$ line. 

Despite our notation with the capital letters, all sublattices of the metals of the AAA and Abb regions in the very vicinity of the aaa regions are weakly correlated. It follows from the fact that the transition between them is continuous while the phases of the aaa regions are noncorrelated.

The less-doped (smaller $|\bar\mu|$) border of the insulators of the abb regions (continuous metal-insulator transition) is perfectly reproduced and analyzed in the MFA as well. However, the further the system from the insulator into the metallic or pinball-liquid phase, the more correlated sublattices become ($Z_\text{a}$ and $Z_\text{b}$ slightly decrease from $1$). The same applies to the b and c sublattices of the AAb, Abb, and Abc regions when moving away from the corresponding insulating phases. Nevertheless, the correlations in the abb regions are weak, the deviations from the MFA results are small and quantitative only, and thus, as discussed in Ref. \cite{AlekseevPRB2025}, the electron-doped (positive $\bar\mu$) conducting phase of the abb region is a pinball liquid, while the hole-doped (negative $\bar\mu$) one is not. 

The low-$zV$ border of the insulators of the abb regions is depicted in Fig. \ref{fig:phase-diagram} (denoted as the mean-field results) and is less relevant because the phases with strong correlations (AAb and Abb regions) turn out to have smaller grand potentials, and thus, there is a discontinuous transition into these regions. Interestingly, when the border of the noncorrelated insulators (mean-field results in Fig. \ref{fig:phase-diagram}) merges with the line of the discontinuous transition with the Abb region, the transition between abb and Abb regions becomes continuous (i.e., there is a tricritical point), see Fig. \ref{fig:aaa_Abb_abb}, $zV=8.05D$.

Finally, it has been found in the MFA that the symmetry breaking between the non-charge-ordered metal to charge-ordered phases is mostly a discontinuous transition which is justified by the approach of the Fermi level to an edge of a band or to a singularity in the density of states \cite{AlekseevPRB2025}. However, a continuous symmetry-breaking transition takes place in the MFA results for small ranges of $\bar\mu$, where the Fermi level falls exactly in between of the compromising points of the band structure during the whole phase transition (see media files from \cite{AlekseevPRB2025}). 
While in the MFA it happens for the model parameters where the phase diagram (Fig. \ref{fig:recap}e) is significantly different to the one of the DMFT (Fig. \ref{fig:phase-diagram}), we have found exactly the same behaviour within the DMFT, where the band-structure analysis is not available anymore. 
Particularly, the continuous transitions between the AAA region and the metals of the Abb and AAb regions (the transition is not visible on the scale of Fig. \ref{fig:phase-diagram}) takes place around the ternary points of the phase diagram---the points where the AAA, Abb, and AAb regions meet. It is very similar to the continuous symmetry-breaking transitions that are located around the ternary point between M$_\text{aaa}$, M$_\text{aab}^1$, and M$_\text{abb}^1$ phases of the MFA phase diagram (Fig. \ref{fig:recap}e). 
The further the system from these ternary points, the sharper the discontinuous phase transitions with the AAA region are, and the more states of the system can be realized within the phase separation only.
See the example of such a continuous phase transition in Fig. \ref{fig:continuous_sym_break}, $zV=3.355D$.
Despite our notation with small letters, the sublattices b of the Abb-region and AAb-region metals can be strongly correlated ($Z_\text{b} < 1$) in the vicinity of such a continuous transition to the AAA region (as in Fig. \ref{fig:M_to_MottPL}).

\subsection{Pinball Liquids}\label{sec:PL}

As discussed in Sec. \ref{sec:phases}, we distinguish two kinds of pinball-liquid phases: Mott-localization-driven PLs and charge-transfer-driven PLs. A Mott-localization-driven PL phase is found within the Abb region only and, strongly asymmetrically, for positive $\bar\mu$ only (for at least as large $U$ as $4D$). It has a continuous transition to the insulator of the Abb region, and characterized by $Z_\text{A} \rightarrow 0$ on metal sides of the transitions with the Abb-region metallic phases (Fig. \ref{fig:M_to_MottPL}).

The PL phase in the abb region for positive $\bar\mu$ is the one found in the MFA \cite{AlekseevPRB2025} as discussed above, and is obviously a charge-transfer-driven PL. 
Contrary to possible expectations, the PL phase that---again asymmetrically---appears in the Abc region is also a charge-transfer-driven PL (for at least as large $U$ as $4D$). Out of two weakly correlated sublattices (b and c) of this Abc-region phase, one is nearly empty, and another is nearly filled. The lattice sites of the one that is nearly empty serve as the pins in the pinball liquid, while the nearly filled sublattice together with strongly correlated sublattice A make up a conducting honeycomb medium.

Finally, the AAb regions contain the charge-transfer-driven PLs. Interestingly, these are much more symmetric with respect to the change of a sign of $\bar\mu$ (or, equivalently, the change of a sign of the hopping amplitude $t$), despite the fact that the AAb regions themselves show strong particle-hole asymmetry. 
One can schematically illustrate the spectral function of the insulators in the AAb regions with three bands (see Fig. \ref{fig:AAb} in the SM \cite{Note1}): lower and upper Hubbard bands whose weights mostly come from the sublattices A, and a band whose weight mostly comes from the sublattice b. The latter is similar to a noninteracting band of a triangular lattice formed by sites of the sublattice b (cf. band formation in the MFA \cite{AlekseevPRB2025}).
In such a picture, the Fermi level of the AAb-region insulator is located between the A-sublattice Hubbard bands and below the b-sublattice band (negative-$\bar\mu$ AAb region) or above the b-sublattice band (positive-$\bar\mu$ AAb region). 
While the actual spectral function is more complicated (see the SM \cite{Note1}), this picture helps to understand why the PL phase is on one side of the insulator and not another: when $|\bar\mu|$ increases, the Fermi level moves away from the b-sublattice band, and eventually b-sublattice sites act as pins in the pinball liquid.
However, we should note that very close to the insulating phases of the AAb regions the analysis of these PLs is complicated by computational problems (see the corresponding PL--insulator transitions in Figs. \ref{fig:Abb_to_AAb} and \ref{fig:continuous_sym_break}), and we cannot guarantee that the phases are always pinball liquids in the vicinity of the insulators.

Note that despite being charge-transfer-driven PLs, the PLs of AAb and Abc regions require Mott physics to exist.

\section{Summary and Final Remarks}\label{sec:summary}

We have presented a rich ground-state phase diagram of the triangular-lattice extended Hubbard model with repulsive nearest-neighbor interaction, charge orders that are commensurate with the $\sqrt{3}\times\sqrt{3}$ supercell, the Mott physics taken into account by means of the DMFT, and the all-encompassing range of chemical potentials with arbitrary charge density.
The phase diagram may further be used in the discussion of the moir\'e lattices, while the quarter-filling results are relevant for the organic conductors.

To identify the phases and phase transitions, a list of order parameters has been analyzed which includes the sublattice occupation numbers, sublattice double occupations, sublattice renormalization constants to estimate the strength of onsite correlations, and sublattice-projected spectral functions at the Fermi level to estimate the charge itineracy, as well as the whole sublattice-projected spectral functions along the real-frequency axis. We have found $5$ pinball-liquid phases and divide them into two types: the charge-transfer-driven PL ($4$ phases) and the Mott-localization-driven PL ($1$ phase). In the former, the charge on "pins" is localized due to a strong intersite repulsion, depside the fact, that $3$ of such PLs still require Mott physics to exist (another one is predicted in the MFA). In the Mott-localization-driven PL the localization comes from the strong onsite correlations. 

During the research the results of the atomic-limit model and mean-field approximation provided useful insight to the physics of the system. We showed that it is possible to set rather good expectations on the DMFT phase diagram based on them while they are much more computationally available. 
The AL results outline the regions of the final DMFT phase diagram and answer the question on what sublattices will be strongly and weakly correlated in each region. 
The MFA provides results for all weakly correlated sublattices, including the whole phase-diagram regions, such as the fully unoccupied, fully occupied, and abb regions (see Fig. \ref{fig:phase-diagram}). 
It is, moreover, correctly predicts the discontinuous symmetry breaking between non-charge-ordered and charge-ordered phases; the existence of their change to the continuous symmetry breakings; the existence of the strong particle-hole asymmetry and some of its aspects; and the existence of nearly occupied charge-ordered metallic phase of the Abb region that purely comes from the finite itineracy on the frustrated geometry and located between the MFA and AL lines. 
The seemingly uninteresting fully occupied and unoccupied phases turn out to be useful to outline the asymmetry of the phase diagram and interesting due to the continuous transition directly to a charge-ordered phase.
The point where the DMFT phase transition between the abb and Abb regions merges with the corresponding MFA transition identifies the tricritical point, i.e., the discontinuous transition becomes continuous. 
Finally, the schematic picture of mean-field bands and Hubbard bands helps to understand the reason the pinball liquids of the AAb regions are on a higher-doped side of the corresponding insulating phases.
An intermediate metallic phase of the ABc region was unexpected in the light of the AL and MFA results. 

Most features appear in the phase diagram asymmetrically for electron and hole doping, thus, the particle-hole asymmetry of a triangular lattice is essential. The electron-doped side is more affected by the Mott physics (e.g., it requires smaller intersite interaction for the strongly correlated charge-ordered phases and contains the Mott-localization-driven PL), which is also the side where the mean-field non-charge-ordered phase has a Van Hove singularity.

A number of discontinuous transitions, primarily between the regions that are colored and outlined by the AL results (Fig. \ref{fig:phase-diagram}), i.e., the existence of jumps of the order parameters including the total charge density, give rise to the phase-separated states, including those for the quarter filling of elections or holes that is characteristic for the organic conductors.

Finally, we should note that in the triangular lattice each lattice site is evenly surrounded by $6$ others in a 2D plane which makes the neglection of the spatial correlations realistic, however, only as far as we are ignoring the existence of the third dimension.
Meanwhile, we should generally expect the existence of magnetic orders across the phase diagram when the onsite interaction dominates over the intersite one, especially for $zV<\frac{U}{2}$. We may also expect that the interesting nearly occupied metallic Abb-region phase should compete with longer-range charge orders that require larger supercells but may be more preferable for low electron or hole densities.

A vast variety of ways to extend this work is available among which: consideration of supercells to take into account the stripe order and longer-range orders, taking into account the magnetic ordering, inclusion of an exchange (Fock) term in the mean-field decoupling of the intersite interaction, introduction of next-nearest-neighbor intersite interactions, attractive interactions, finite temperatures, taking into account spatial correlations by means of an extended DMFT, and more, yielding more and more complex phase diagrams. One should note however that already the current result required considerable computational time and associated with various convergence problems of the self-consistency algorithm.


%

\clearpage

\onecolumngrid

\linespread{1.3}
\selectfont

\begin{center}
  \textbf{\Large Supplemental Material}\\[.2cm]
  \textbf{\large Charge order on a triangular lattice with Mott physics and arbitrary charge density}\\[.2cm]
  Aleksey Alekseev,$^{1}$ Agnieszka Cichy,$^{1,2}$ Konrad Jerzy Kapcia,$^{1}$ \\ [.2cm]
  {\itshape
  	$^{1}$\mbox{Institute of Spintronics and Quantum Information, Faculty of Physics and Astronomy}, \mbox{Adam Mickiewicz University in Pozna\'n}, 
    Uniwersytetu Pozna\'{n}skiego 2, PL-61614 Pozna\'{n}, Poland\\
	$^{2}$Institut f\"{u}r Physik, Johannes Gutenberg-Universit\"{a}t Mainz, Staudingerweg 9, D-55099 Mainz, Germany\\
	}
(Dated: \today)
\\[1cm]
\end{center}

\setcounter{equation}{0}
\renewcommand{\theequation}{S\arabic{equation}}
\setcounter{figure}{0}
\renewcommand{\thefigure}{S\arabic{figure}}
\setcounter{section}{0}
\renewcommand{\thesection}{S\arabic{section}}
\setcounter{table}{0}
\renewcommand{\thetable}{S\arabic{table}}
\setcounter{page}{1}

\linespread{1.4}
\selectfont

\begin{figure}[b]
    \centering
    \includegraphics[width=1\linewidth]{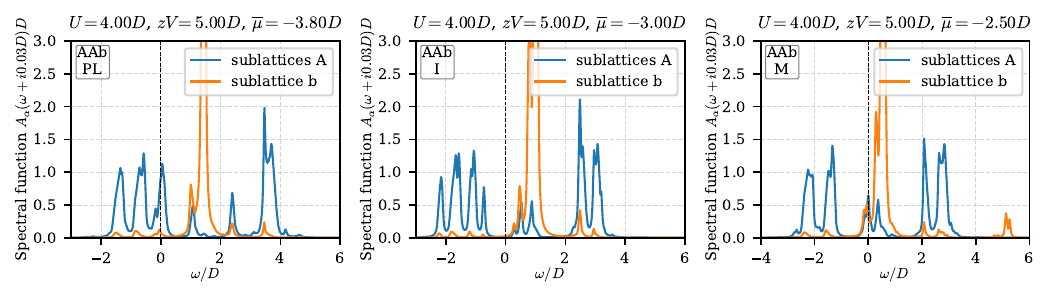}
    \caption{Sublattice-projected spectral functions of the phases of the AAb region, from left to right: pinball-liquid (PL), insulating (I), and metallic (M) phase. 
    The model parameters as labeled.}
    \label{fig:AAb}
\end{figure}

\begin{figure}[b]
    \centering
    \includegraphics[width=1\linewidth]{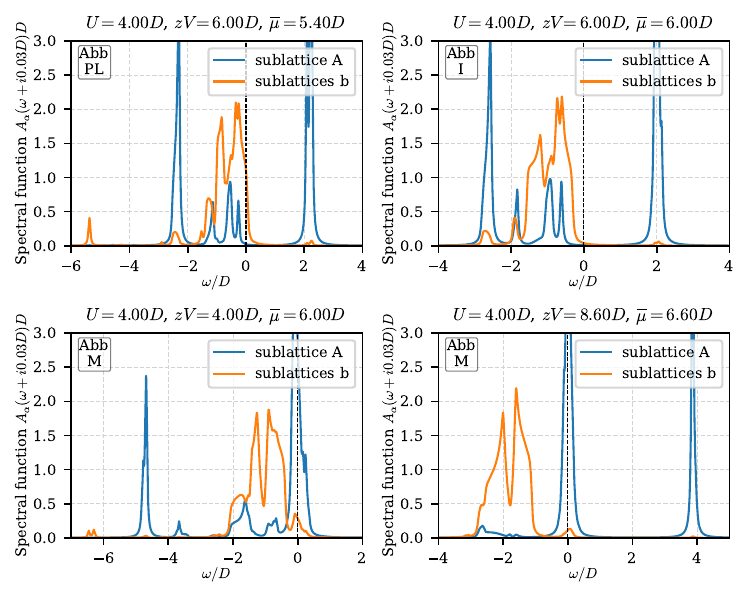}
    \caption{Sublattice-projected spectral functions of the phases of the Abb region (pinball-liquid (PL), insulating (I), and two metallic (M) phases). 
    The model parameters as labeled. 
    The metallic phase on the bottom left panel is shown for such model parameters where it is already far from the continuous transition to the fully occupied phase, and is about to transition to the Abb-region insulating phase.
    The metallic phase on the bottom right panel is shown for such model parameters where it is about to transition to the abb-region insulating phase.
    }
    \label{fig:Abb}
\end{figure}

\begin{figure}[t]
    \centering
    \includegraphics[width=1\linewidth]{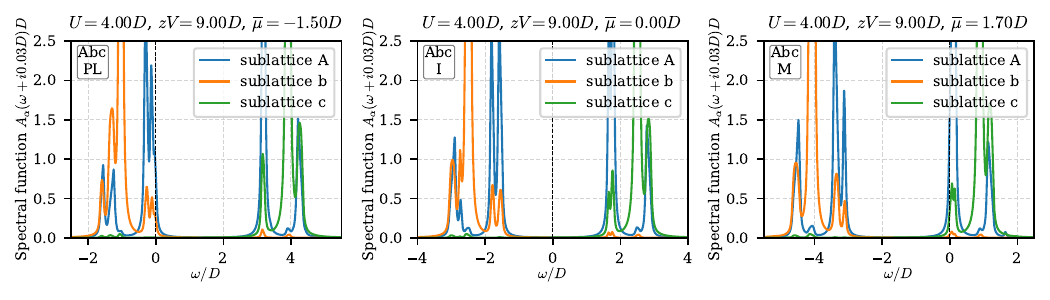}
    \caption{Sublattice-projected spectral functions of the phases of the Abc region (pinball-liquid (PL), insulating (I), and metallic (M) phase). 
    The model parameters as labeled.}
    \label{fig:Abc}
\end{figure}

\begin{figure}[t]
        \centering
        \includegraphics[width=1\linewidth]{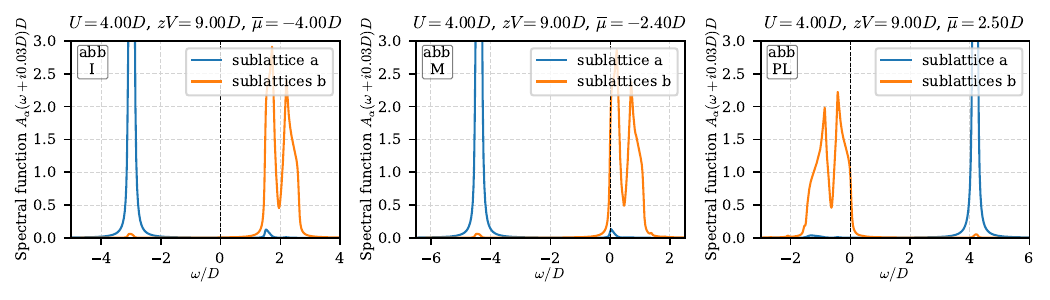}
        \caption{Sublattice-projected spectral functions of the phases of the abb region (insulating (I), metallic (M), and pinball-liquid (PL) phase), that reproduce the mean-field results (Phys. Rev. B \textbf{112}, 115155 (2025)).
        The model parameters as labeled.}
        \label{fig:abb}
\end{figure}

In this Supplemental Material we show a few representative sublattice-projected spectral functions of the phases of
\begin{itemize}
    \item the AAb region -- Fig. \ref{fig:AAb}, 
    \item the Abb region -- Fig. \ref{fig:Abb},
    \item the Abc region -- Fig. \ref{fig:Abc}, 
    \item and the abb region -- Fig. \ref{fig:abb}.
\end{itemize}
In the figures, the vertical black dashed line indicate the Fermi level ($\omega=0D$).
For the AAb region, $A_\text{A}(z) = A_1(z) = A_2(z)$ and $A_\text{b}(z) = A_3(z)$, and correspondingly for each region. 
Similarly as in the main text, we use capital letters (A, B) to show that a sublattice is strongly correlated and small letters (a, b, c) otherwise.

$\quad$

Note that for the pinball-liquid phases the $A_\alpha(\omega+i0.03D)D$, where $\alpha$ is a sublattice of ``pins'', is far from $0$ because the imaginary part of the frequency ($i0.03D$) is large. 
The parameter has such a large value to make the spectral functions representative instead of having a set of sharp peaks (see, e.g., Rev. Mod. Phys. \textbf{68}, 13 (1996)).

$\quad$

The most of the spectral weight of the strongly correlated sublattices (sublattices A of the regions AAb, Abb, and Abc) is located in two frequency regions that can be identified as lower and upper Hubbard bands. 
The most of the spectral weight of the weakly correlated sublattices (sublattices b of the regions AAb and Abb; sublattices b and c of the region Abc; sublattices a and b of the region abb) is located withing a single band. 
In the regions Abb and abb, the spectral function of the sublattices b is similar to the noninteracting honeycomb density of states, while the spectral functions of the abb-region phases reproduce those obtained in the mean-field approximation (cf. Phys. Rev. B \textbf{112}, 115155 (2025)).
The relative location of the Hubbard and weakly correlated bands, together with the location of the Fermi level, depends on the place in the phase diagram.

\clearpage

\end{document}